# PREDICTIVE THEORY OF NEUTRINO MASSES

Olakanmi F. Akinto *,†,‡ and Farida Tahir *,#

*Department of Physics, COMSATS University Islamabad, Pakistan

†Department of Physics, National Mathematical Centre, Sheda-Kwali Abuja, Nigeria

‡pheligenius@yahoo.com

#farida_tahir@comsats.edu.pk

In our recent paper [1] we formulated a predictive theory of neutrino masses by considering the interaction between the infrared sector of the effective theory of quantum gravity and the standard model fields. This allowed us to calculate, for the first time in the history of neutrino physics, the absolute scale of neutrino masses. From this theoretical framework, we obtained quantum-gravitational couplings/effective Majorana dimensionless couplings from the spherically symmetric vacuum solutions arising from the Bose-Einstein statistical modification to gravitation. In the present paper, we show that the same solutions can be obtained directly from the quantum interpretation of gravitational radiation arising from the thermodynamic modification to gravitation. Within this theoretical scheme, we show that the single-field inflationary model, GUTs, dark energy and matter-independent gravitational field of vacuum are all connected to the neutrino mass model.

**Keywords**: Neutrino masses, GUTs, quantum gravity

## I. INTRODUCTION

Neutrinos have been around, literally, since the beginning of time. In the sweltering moments following the Big-Bang epoch, neutrinos were among the first sub-atomic particles to emerge from this primordial fireball. Neutrinos are transients, interacting only through the weak force and gravity thereby tracing long, lonely trajectories across the universe. This property makes neutrinos the most eloquent footprint of a possible quantum-classical nature of gravity, representing a unique probe for investigating physics at all distance scales [1-2].

However, the question of how neutrino masses arise and their absolute values have not been answered conclusively: In the standard model

(SM) of particle physics, neutrinos are predicted as massless and electrically neutral (sub-atomic) fermions, but the experimental results of all the neutrino-oscillation experiments have shown, convincingly, that neutrinos do, indeed, have a vanishingly small mass obtained from squared-mass differences $\Delta m_{ji}^2 = m_j^2 - m_i^2$ [3]. Precision measurements of the flux of solar neutrinos revealed that fewer electron-type neutrinos arrived at the Earth than predicted (about one-third of the predicted value). This problem was definitely resolved by the SNO experiment in the 2000s. Eventually solar neutrino data imply that electron-neutrinos are linear superpositions of at least two neutrino mass-eigenstates. In this case the difference between the neutrino squared-masses is of order $\Delta m_{21}^2 = m_2^2 - m_1^2 \sim 10^{-4} eV^2$. Where $m_1$ and $m_2$ are the nonzero masses $(i.e., m_2 > m_1 \neq 0)$ of two different neutrino mass-eigenstates, labelled $\mathcal{V}_1$ and $\mathcal{V}_2$. Similarly, the precision measurements of the flux of atmospheric neutrinos also revealed that about one-third of the muon-type neutrinos survived passage through the Earth than predicted. The solution to this atmospheric neutrino problem was the realization that muon-neutrinos are also linear superpositions of at least two neutrino mass-eigenstates, where $m_3 \gg m_1 \neq 0$. Here the squared-mass difference is of order $\Delta m_{31}^2 = m_3^2 - m_1^2 \sim 10^{-3} eV^2$ and $m_3$ is the mass of the third distinct neutrino mass-eigenstate, labeled $\mathcal{V}_3$. Nufit Collaboration and other world neutrino oscillation data have also measured, with very good precision (but with calculative assumption that the minimum neutrino mass is zero ( $i.e., m_1 = 0$)), the two independent neutrino squared-mass differences as $\Delta m_{21}^2 = (7.42^{+0.21}_{-0.20}) \times 10^{-5} eV^2$ and $\Delta m_{31}^2 = (2.517^{+0.026}_{-0.028}) \times 10^{-3} eV^2$ for normal mass ordering (NMO). In this case the mass-sum is given as $\sum m_\mathcal{V} = \sqrt{\Delta m_{21}^2} + \sqrt{\Delta m_{31}^2} \simeq 0.0588 eV$ [4].

Unfortunately, neutrino oscillations are not sensitive to the absolute neutrino masses $m_i$ and $m_1 = 0$ (for NMO) is an artefact of calculative assumption used in neutrino-physics community which is in strong tension with the Nature's modus operandi as we have shown in [1]. Hence further information about the absolute scale of neutrino masses must be investigated from non-oscillation approaches, using beta decay, neutrinoless double beta decay, or cosmological observations [5]. Presently, there are two known natural explanations of the smallness of neutrino masses with respect to those of the other SM elementary particles: (i) the see-saw mechanism [6] and (ii) the mechanism of effective generation of a Majorana/Weinberg mass term by physics beyond the SM (BSM). In both cases massive neutrinos are Majorana particles, leading to the possibility of observing neutrinoless double beta decay [3].

After decades of intense experimental and theoretical research, we still do not know the mechanism that leads to nonzero neutrino masses $m_i$. The purpose of [1] and this review is to discuss our formulation of neutrino mass model constructed from the gravity-embedded Weinberg operator. This idea is based on the fact that neutrinos only participate in weak and gravitational interactions. Hence one could construct a simple theory of neutrino masses by considering the interaction between the infrared sector of the effective theory of quantum gravity and the SM fields. Effectively, these masses are generated by the gravity-embedded 5-dimensional Weinberg operator [1]:

$$M_{\mu\nu} = g_{\mu\nu} \frac{v_{Ew}^2}{\Lambda_{GUT}}. \quad [1]$$

Where $M_{\mu\nu} = m_\mu \delta_{\mu\nu}$ ($\mu, \nu = 0,1,2,3 \text{ or } 1,2,3,4$), and $\delta_{\mu\nu} = \begin{Bmatrix} 0 \to if\ \mu \neq \nu \\ 1 \to if\ \mu = \nu \end{Bmatrix}$ [3]. Since $\mu = \nu$ for both Majorana neutrino masses $M_{\mu\nu}$ and gravity, Eq. (1) reduces to

$$m_\mu = g_{\mu\nu} \frac{v_{Ew}^2}{\Lambda_{GUT}}. \quad [2]$$

Where $\mathcal{V}_{Ew}(= 246 GeV)$ is the electroweak vacuum expectation value of Higgs field and $\Lambda_{GUT}$ is the spontaneous symmetry-breaking mass-scale of Grand Unified Theories (GUTs).

Equation 2 is in agreement with that in [7]. The authors derived this formula ($p_\mu^{(k)} \frac{ds}{dx^\mu} = m_k g_{\mu\nu}$) by considering the propagation of neutrinos in a gravitational field of a non-rotating spherically symmetric object, which is described by the Schwarzschild metric. Although they arrived at the correct result, their derivation is inconclusive because the values of their mass term ($m_k$) and gravitational field ($g_{\mu\nu}$) were not clearly defined. Here our theoretical intervention is two-fold: (i) to show that the derivation in [7] is compatible with the Nature and (ii) to solve Eq. (2) completely within the contextual scheme of GUTs and general radiative solutions of the exact Einstein field equations.

## II. GRAND UNIFIED MODELS, SINGLE-FIELD INFLATIONARY MODEL AND THEORY OF NEUTRINO MASSES

GUTs are one of the most interesting high-energy completions of the SM because they provide a rich and powerful group-theoretic framework able to solve many BSM problems, such as hot Big-Bang cosmology.

The standard model of hot Big-Bang cosmology relies on the assumption that: as the time (t) approaches zero, the temperature T approaches infinity. That is, as $t \to 0$, $T \to \infty$ (where Energy $(E) = Mass(M) = Temperature\ (T)$, in natural units $k_B = \hbar = c = 1$). Since the behaviour of the universe during the first fraction of a second $t \lesssim 10^{-44}s$ ( and $T \gtrsim 1.22 \times 10^{19} GeV$) after the Big-Bang can only be a matter for conjecture, it is plausible to construct the theory of hot Big-Bang cosmology at some temperature $T_0 = 10^{17} GeV$ based on GUTs and special solutions of the exact Einstein field equations. At this temperature, the description of the universe is taken as a set of initial conditions [8]. This characteristic temperature (i.e., $T_0 = 10^{17} GeV$) is the unification temperature-scale of grand unified models [9].

Within the Big-Bang cosmology it is assumed that the universe expands and cools. During this cooling epoch, the universe passes through some critical temperatures corresponding to characteristic energy scales. These transitions are connected with symmetry breakings and the inflationary expansion of the early universe [8]. A general picture thus emerges of a grand unified model which begins with a simple gauge group $\mathcal{G}_{GUT}$ and a valid symmetry at the highest energy $M_U = 10^{17} GeV$. As the energy is considerably lowered (due to inflation), the theory undergoes a two-step hierarchy of spontaneous symmetry breaking into successive subgroups:

$$\mathcal{G}_{GUT} \xrightarrow{\Lambda_{GUT}} \mathcal{G}_{SM} \xrightarrow{\mathcal{V}_{EW}} \mathcal{G}_{EM} \qquad [3]$$

Where $\mathcal{G}_{SM} = SU(3)_C \times SU(2)_L \times U(1)_Y$ and $\mathcal{G}_{EM} = SU(3)_C \times U(1)_Q$. In the type-A 331 model, which is the simplest non-supersymmetric model of the type $\mathcal{G}_{331} \xrightarrow{\Lambda_{GUT}} \mathcal{G}_{SM}$, the symmetry breaking occurs at the minimum energy scale $\Lambda_{GUT} = 1.63 \times 10^{16} GeV$ for $M_U = 10^{17} GeV$ [9]. (It is worth noting here that 331 model can admit any simple groups like $E_6, SU(5), SU(7), SO(10)$, etc. since it is an intermediate gauge group between the SM and the scale of unification of the three non-gravitational interactions). This value $(i.e., \Lambda_{GUT} = 1.63 \times 10^{16} GeV)$ is consistent with the value predicted $(i.e., \Lambda_{GUT} \approx 2.0 \times 10^{16} GeV)$ by many minimal supersymmetric GUTs and string-derived $G(224)$ [10-12].

We now have a hierarchy of two vacuum expectation values $\mathcal{V}_{EW}$ and $\Lambda_{GUT}$ regulated by Majorana neutrino mass (see Eqs 1 and 2):

$$\langle m \rangle_{eff} \equiv \frac{\mathcal{V}_{EW}^2}{\Lambda_{GUT}} = 3.7126 meV. \qquad [4]$$

(This value is to be compared with the effective Majorana neutrino mass $1.2 meV \lesssim \langle m_{ee} \rangle \lesssim 4 meV$ given within the range of 5-sigma in [13]).

Since GUTs play an important role in the inflationary dynamics of the early universe, their imprints could be found in the cosmic microwave background (CMB) observations by the Planck satellite [14]. Hence for the cosmic inflation to be compatible with the symmetry-breaking scheme given in Eq. (3), the vacuum energy $(\Lambda_{inf})$ responsible for the inflationary expansion of the universe must be equal to $\Lambda_{GUT}$: For single-field inflationary model, the observed amplitude of CMB anisotropies implies that [14]

$$\Lambda_{inf} = 2.2 \times 10^{16} GeV \left(\frac{r}{0.2}\right)^{1/4}. \quad [5]$$

Where $r$ is the tensor-to-scalar ratio. Its peak value has been measured from the Planck temperature measurement, BICEP2 and keck instruments to be $r = 0.06$ [15]. This same upper value had been previously obtained when the spectral index $n_s = 0.98$ in [16]. With this value of $r(= 0.06)$, Eq. (5) reduces to

$$\Lambda_{GUT} = \Lambda_{inf} = 1.63 \times 10^{16} GeV \quad [6]$$

Hence Eq. (3) can be regarded as the particle-physics mechanism responsible for the cosmic inflation.

### A. Cosmology and Gravitational couplings

A very different way to get information on the absolute scale of neutrino masses is from the study of the CMB radiation spectrum (with thermodynamic temperature $T_\gamma^0$) as well as the study of the large scale structure in the universe. In the standard model of cosmology, neutrinos are predicted to be relics of the hot Big-Bang. Here, the measurements of the relic abundance of light elements are consistent with the existence of a cosmic neutrino background (C$\mathcal{V}$B) with thermodynamic temperature $T_\nu^0$.

The thermodynamic temperatures of CMB and C$\mathcal{V}$B are related by [3, 17]

$$T_\gamma^0 = \left(\frac{11}{4}\right)^{1/3} T_\nu^0. \quad [7]$$

Where $T_\nu^0 \sim 2 \times 10^{-4} eV$. As the universe expands and the relic neutrino background cools, the behaviour of neutrinos changes from that of ultrarelativistic relics (radiation) to that of nonrelativistic species (matter) as long as $\sum m_\mathcal{V} \gg T_\nu^0$ [4]. This radiation-matter transition is governed by $E = T = M$ in natural units ($k_B = \hbar = c = 1$). This transition should leave an imprint in the large-scale structure of the universe in such a way that precision measurements could provide nontrivial information on the neutrino masses. In this case the relic neutrino background is best described as a homogeneous mixture of the neutrino mass-eigenstates $\mathcal{V}_i$, where cosmic surveys are expected to provide information on the individual neutrino masses $m_i$. By combing $M = T$ with the spherically symmetric vacuum solutions of Einstein field equations in their isotropic form, and using Majorana conditions ($\mathcal{V}_L = \mathcal{V}_L^C$, $\Theta = n\pi$ & $sin\theta = 0$, $and\ r = \sqrt{3}$) [1], we get

$$g_{00} = \frac{\left(1-\frac{T}{2\sqrt{3}}\right)^2}{\left(1+\frac{T}{2\sqrt{3}}\right)^2}, \quad [8]$$

$$g_{11} = \left(1+\frac{T}{2\sqrt{3}}\right)^4, \quad [9]$$

$$g_{22} = 3 \times \left(1+\frac{T}{2\sqrt{3}}\right)^4, \quad [10]$$

$$g_{33} = 0. \quad [11]$$

Where T is the thermodynamic temperature.

To further analyze Eqs 8-11, it is convenient to put the thermodynamic temperature T in dimensionless form (by using Eq. (7)) since the radial distance $r$ is dimensionless. Defining

$$T \to \frac{T_\gamma^0}{T_\nu^0} = \left(\frac{11}{4}\right)^{\frac{1}{3}} = 1.4010. \quad [12]$$

From Eqs 8-12, the gravitational couplings are calculated as

$$\left.\begin{array}{l} g_{00} = 0.1798, \\ g_{11} = 3.8905, \\ g_{22} = 11.6715, \\ g_{33} = 0. \end{array}\right\} \quad [13]$$

Equation 13, which is the matter-independent gravitational field of vacuum, is to be compared with the values of $g_{\mu\nu}$ in [1].

From Eqs (2) and (13), the individual neutrino masses are calculated as

$$\left.\begin{array}{l} m_1 = 0.6675\, meV, \\ m_2 = 14.4439\, meV, \\ m_3 = 43.3316\, meV, \\ m_4 = 0\, meV. \end{array}\right\} \quad [14]$$

And the mass-sum for the three active neutrinos is given as

$$\sum_{i=1}^{3} m_i = m_1 + m_2 + m_3 = 0.0584\, eV. \quad [15]$$

This value is to be compared with the minimum value, $\sum m_\nu \simeq 0.0588\, eV$, obtained from the neutrino-oscillation experiments [3-4]. Eq. (15) is also consistent with the $0.058\, eV \lesssim \sum m_\nu \lesssim 0.062\, eV$ obtained within the range of 5-sigma in [13].

## B. Atmospheric Neutrino and Solar Neutrino Squared-Mass Differences

(i) The atmospheric neutrino squared-mass difference is calculated (from Eq. (14)) as

$$\Delta m_A^2 = \Delta m_{31}^2 = m_3^2 - m_1^2 = 1.88 \times 10^{-3}\, eV^2 \quad [16]$$

Equation 16 is consistent with the measurement of atmospheric neutrinos and antineutrinos in the MINOS far detector, where $\Delta m_{31}^2 = 1.9 \times 10^{-3}\, eV^2$ for 90% single-parameter confidence intervals [18].

(ii) The solar neutrino squared-mass difference is calculated (from Eq. (14)) as

$$\Delta m_\odot^2 = \Delta m_{21}^2 = m_2^2 - m_1^2 = 2.08 \times 10^{-4}\, eV^2 \quad [17]$$

Equation 17 is consistent with all solar neutrino oscillation experiments with KamLAND data measured as $\Delta m_{Sol}^2 = \Delta m_{21}^2 \gtrsim 2.0 \times 10^{-4} eV^2$ at 99.73% C.L.[19].

Thus the mass-sum of the three active neutrinos can be expressed in terms of the first part of Eq. (14), Eqs 16 and 17 as

$$\sum m_\nu = m_1 + \sqrt{\Delta m_{21}^2 + m_1^2} + \sqrt{\Delta m_{31}^2 + m_1^2} = 0.0585 eV \quad [18]$$

### C. Cosmological Constant/Dark Energy

If truly the vacuum energy owes its existence to the asymmetry/broken symmetry of the cosmos, then the effective Majorana neutrino mass regulates the value of vacuum energy as $\epsilon_{vac} = \langle m \rangle_{eff}$. Eq. (4) is to be compared with $\epsilon_{vac} \approx 3.6 meV$ in [20]. Hence the proper vacuum energy density (or dark energy), responsible for the accelerated expansion of the universe, is given as [1]:

$$\rho_{vac} = T_{00}^{vac} = g_{00}[\langle m \rangle_{eff}]^4 = (2.42 \times 10^{-3} eV)^4 \quad [19]$$

This value is to be compared with two different theoretical approaches: (i) from the resummed quantum gravity in the context of asymptotic safety, we have $\rho_\Lambda \simeq (2.4 \times 10^{-3} eV)^4$ [21]. And (ii) from low energy quantum gravity in the context of QCD Veneziano ghost, we have

$\rho_\Lambda \simeq c_{grav} \times (3.6 \times 10^{-3} eV)^4$ [20], where $c_{grav}$ is a dimensionless gravitational coupling. These three different theoretical approaches are consistent with the observed value $\rho_\Lambda \simeq ((2.37 \pm 0.05) \times 10^{-3} eV)^4$ [21].

### III. CONCLUDING REMARKS

We have explored a predictive neutrino mass model with a gravity-embedded Weinberg operator, in which we have generated the neutrino masses through the interaction between the infrared sector of the effective theory of quantum gravity and the standard model fields. Embedding gravity into a two-step spontaneous symmetry breaking mechanism of grand unified models, we predict three active neutrinos with nonzero masses $m_{1,2,3} \neq 0$ and one inactive neutrino with zero mass $m_4 = 0$. This remarkable yet simple idea owes its unique existence to the quantum interpretation of gravitational radiation arising: from the Bose-Einstein statistical modification to gravitation in [1], and to the spherically symmetric vacuum solutions arising from the thermodynamic modification to gravitation as established in this paper. These two different theoretical approaches only point to one conclusion that neutrinos produced in a weak interaction (governed by extended electroweak theory) must propagate in a gravitational field in order to manifest their oscillation property because the squared-mass differences $\Delta m_{ji}^2$ depend only on the different values of $g_{\mu\nu}$. We argue that if the values of $g_{\mu\nu}$ were to be the same, $\Delta m_{ji}^2$ would vanish completely, and thus no oscillation would have been observed. Hence neutrino-oscillation phenomenon is a direct consequence of gravity-embedded 5-dimensional Weinberg operator. If Equation 2 is correct, one can make a robust prediction that: Neutrinos are Majorana particles and lepton number is not an exact symmetry of Nature. Determining the nature of the neutrinos (whether Dirac or Majorana particles) has received a high priority in particle physics. The most potent probes of the nature of the neutrino are tests of lepton number violation/conservation, especially searches for neutrinoless double-beta decay. Another important prediction of the theoretical scheme presented in this paper states that dark energy/proper vacuum energy density owes its existence to the two-way symmetric breaking of non-supersymmetric GUT and the matter-independent gravitational field of vacuum.